\documentclass[%
 reprint,
 floatfix,
superscriptaddress,
 longbibliography,
 amsmath,amssymb,
 aps,
]{revtex4-1}

\makeatletter
\def\set@curr@file#1{%
 \begingroup
 \escapechar\m@ne
 \xdef\@curr@file{\expandafter\string\csname #1\endcsname}%
 \endgroup
}
\def\quote@name#1{"\quote@@name#1\@gobble""}
\def\quote@@name#1"{#1\quote@@name}
\def\unquote@name#1{\quote@@name#1\@gobble"}
\makeatother

\usepackage{graphicx}
\graphicspath{ {./images/} }
\usepackage{dcolumn}
\usepackage{bm}
\usepackage{xfrac}
\usepackage{siunitx}
\usepackage{amsmath}
\usepackage[utf8]{inputenc}
\usepackage[T1]{fontenc}
\usepackage[export]{adjustbox}
\usepackage[colorinlistoftodos,prependcaption,textsize=tiny]{todonotes}
\usepackage{gensymb}
\usepackage{upgreek}

\presetkeys%
 {todonotes}%
 {inline, backgroundcolor=yellow}{}

\begin{document}

\title{Transport-Induced-Charge Electroosmosis}

%

\author{Wei-Lun Hsu}
 \email{Corresponding author: wlhsu@thml.t.u-tokyo.ac.jp}
\author{Zhixuan Wang}
\affiliation{Department of Mechanical Engineering, The University of Tokyo, Hongo, Tokyo 113-8656, Japan}
\author{Hirofumi Daiguji} 
\affiliation{Department of Mechanical Engineering, The University of Tokyo, Hongo, Tokyo 113-8656, Japan}


\begin{abstract}
We report theoretical analysis of transport-induced-charge electroosmosis (TICEO) in a nanopore due to the presence of a local electric field and a conductivity gradient. TICEO shares a similar characteristic with classical induced-charge electroosmosis (ICEO) that the mean velocity $\overline{v_\text{TIC}}$ is proportional to the square of the applied electric potential difference $\Delta\psi$, $i.e.$, $\overline{v_\text{TIC}} \propto (\Delta\psi)^2$, appropriate for alternating current (AC) pumping applications. In contrast to ICEO which is primarily used in microfluidics, TICEO does not require metallic or dielectric patches and is thus suitable for nanopore pumping, providing new opportunities for AC nanopore applications.

\end{abstract}

\maketitle

 
In the past two decade, owing to the advent of nanometer-sized pore drilling methods using transmission electron microscopy and electron-beam lithography,~\cite{Storm2003, Verschueren2018} the dimensions of interest in electrokinetics have gradually shifted from the microscale to the nanoscale~\cite{Alizaheh2021, Lee2018}.  
Especially, the emergence of low aspect ratio (pore length/pore diameter) nanopores offers tremendous promises to high precision biosensing for rapid medical diagnosis and high power density energy conversion~\cite{Tsutsui2012, Daiguji2004}. 
Along with the electric field, a salt gradient usually coexists, which is either intentionally imposed or locally induced across these nanopores~\cite{Hsu2016, Hatlo2011}. For instance, a salt concentration bias was added across nanopores to enhance DNA sensing efficiency~\cite{Wanunu2010}. On the other hand, a local concentration difference can arise across an ion selective membrane due to ion concentration polarization (ICP)~\cite{Zangle2010}. 
Hence, when applying an electric field, a concurrent solute concentration difference inevitably appears in most nanopore systems.

In this regard, a nonuniform electric field caused by the salt gradient may give rise to local ion separation in electrolyte solutions~\cite{MacGrillivray1968, Dickinson2011}. He $et$~$al.$ indicated that ionic charges can be induced near the entrance of a nanopore, which might facilitate the capture of DNA molecules for resistive pulse sensing~\cite{He2013}. In nanochannels, Zhu $et$~$al.$ pointed out locally induced charges yield inverse screening effects by coions, giving rise to local vortices~\cite{Zhu2016}. Hsu $et$~$al.$ identified that the sign of induced charges reverses with the reversal of the external field direction, resulting in a unidirectional flow from the high concentration side to the low concentration reservoir in a neutral nanopore in spite of the direction of the imposed electric field~\cite{Hsu2018}. Built upon this, they proposed a concept of alternating current (AC) nanopores resistive pulse sensing.  
To distinguish from classical induced-charge electroosmosis (ICEO) generated around a polarized metallic or dielectric material~\cite{Davidson2014,Bazant2004}, the term "transport-induced-charge electroosmosis (TICEO)" was first introduced by Hsu $et$~$al.$ to describe the electroosmotic flow originated from induced charges in an electrolyte solution due to the presence of a local electric field and a solute concentration gradient~\cite{Hsu2018}. 
Wang $et$~$al.$ investigated thermal effects on transport-induced-charge (TIC) phenomena~\cite{Wang2020}. 

\begin{figure}[t!]
	\includegraphics[width=\columnwidth]{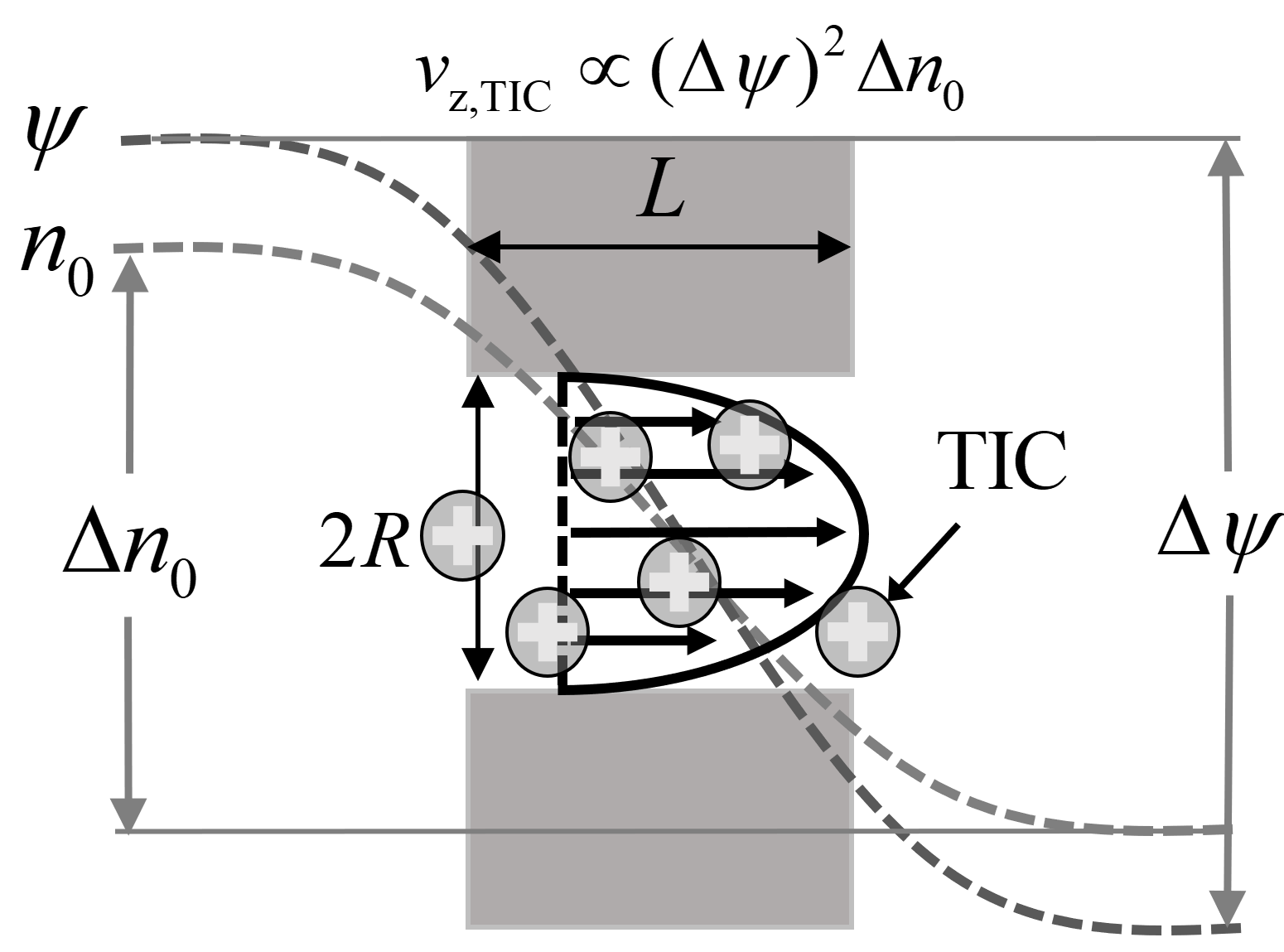}
	\caption{\label{fig1} Schematic illustration of transport-induced-charge (TIC) electroosmotic velocity $v_{z,\text{TIC}}$ distribution in a nanopore as an electric potential bias $\Delta\psi$ and a salt concentration difference $\Delta n_0$ are concurrently imposed. $L$ and $R$ are the pore length and pore radius, respectively.}
\end{figure}

In this Letter, we analytically verify TIC phenomena in a nanopore in the presence of an external electric field and a salt concentration difference, as schematically illustrated in Fig. 1. It is worth highlighting that in addition to the electric field the required condition for TIC phenomena can not only be offered by a salt concentration gradient, but also a temperature gradient (known as electrothermal effects~\cite{Boika2008}), or connecting two solutions with asymmetric viscosity or permittivity as long as a nonuniform conductivity/permittivity field is generated~\cite{Wang2020}.  
Herein, we aim to elucidate the time scale, length scale and typical voltage ranges for the occurrence of TICEO in a solute concentration biased nanopore. 

Ion transport in a nanopore obeys the conservation of current derived from the Nernst-Planck equations of ions as the following:
\begin{eqnarray}
 \frac{\partial\rho_\text{e}}{\partial t}=-\nabla\cdot\textbf{J},
&&\label{eq:one}
\end{eqnarray}
where $\rho_\text{e}$, $t$ and $\textbf{J}$ denote the space charge density, time and current density, respectively. For highly concentrated electrolyte solutions in which (i) the external concentration difference $\Delta{n_0}$ is much lower that the average bulk concentration $n_0$ and (ii) the Debye length $\lambda_\text{D}$ is thin compared with the pore radius $R$ rendering the Dukhin number $<<$ 1. Therefore, the current density can be approximated by the solution conductivity $\sigma_\text{c}$, $i.e.$, $\textbf{J}\cong\sigma_\text{c}\textbf{E}$, where $\textbf{E}$ is the electric field, implying that both the diffusion current and advection current are negligible compared with the magnitude of the conduction current. From Coulomb's law and Gauss's theorem, the relation between $\textbf{E}$ and $\rho_\text{e}$ can be described by the following equation when considering the permittivity of the solution (being the product of the permittivity at vacuum $\varepsilon_0$ and dielectric constant $\varepsilon$):
\begin{eqnarray}
 \nabla\cdot(\varepsilon_0\varepsilon\textbf{E})=\rho_\text{e}.
 &&\label{eq:two}
\end{eqnarray}
Eq.~(2) becomes Poisson's equation as long as expressing \textbf{E} in terms of the electric potential $\psi$ ($i.e.$, $\textbf{E}=-\nabla \psi$). 

\begin{figure}[th!]
	\includegraphics[width=\columnwidth]{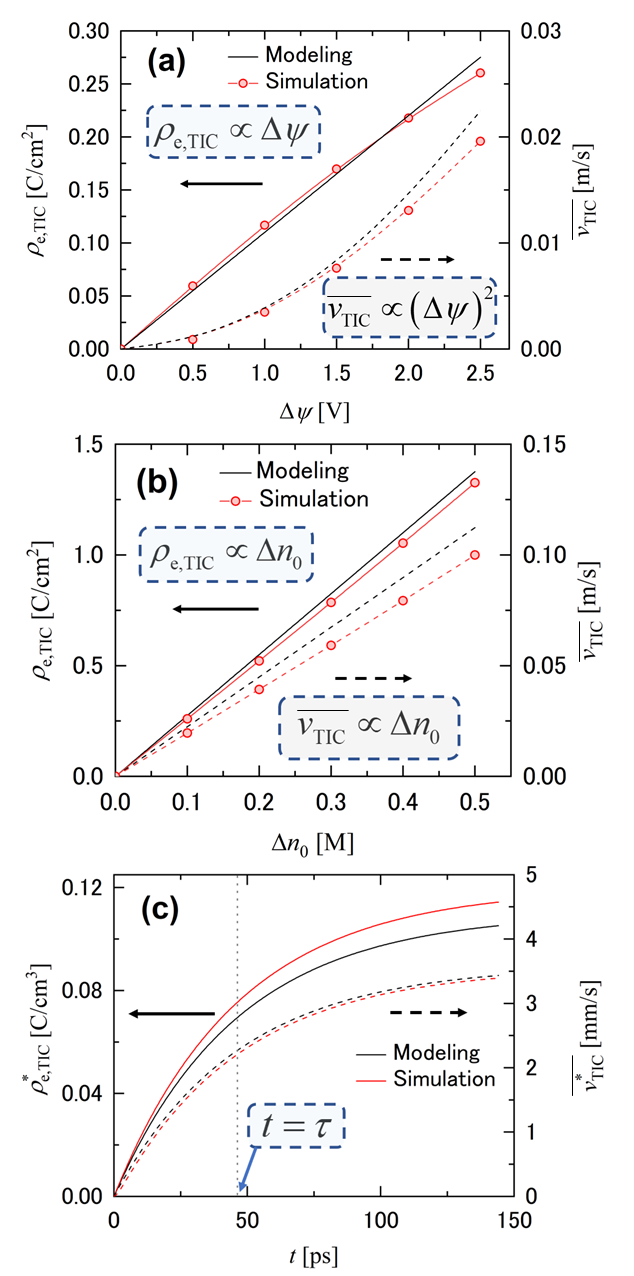}
	\caption{Comparisons between the analytical solutions (black curves) and numerical simulation results (red curves) for the steady-state/transient TIC density $\rho_\text{e,TIC}$/$\rho_\text{e,TIC}^{*}(t)$ (solid curves) and mean TIC electroosmotic  velocity $\overline{v_\text{TIC}}$/$\overline{v_\text{TIC}^{*}}(t)$ (dashed curves) in a nanopore of length $L$ = 20 nm and radius $R$ = 2.5 nm:~(a) $\Delta{n_0}$ = 0.1 M for different $\Delta\psi$, (b) $\Delta\psi$ = 2.5 V for different $\Delta{n_0}$ and (c) transient variations at $\Delta{n_0}$ = 0.1 M and $\Delta\psi$ = 2.5 V, where $\tau$ is a characteristic time $\cong$ 46.3 ps for $n_0$ = 1 M. (The simulated charge density is derived from the average value in the middle of the pore.) \label{fig2}} 
\end{figure}

$\emph{Steady state \textemdash}$ Substituting Eq.~(2) into Eq.~(1) and letting $\partial\rho_\text{e}/\partial t=0$, the steady-state induce charge distribution in the solution is:    
\begin{eqnarray}
\rho_\text{e}=\varepsilon_0\varepsilon \textbf{E}\cdot\bigg(\frac{\nabla\varepsilon}{\varepsilon}-\frac{\nabla\sigma_\text{c}}{\sigma_\text{c}}\bigg).
&&\label{eq:three}
\end{eqnarray}
Thus, by considering (i) linear distributions of the salt concentration and electric potential in a nanopore filled with an isothermal solution ($i.e.$, when Joule heating is negligible) without a permittivity gradient ($i.e., \varepsilon_0\nabla\varepsilon=\textbf{0}$) and (ii) uniform induced charges in the nanopore, the steady-state TIC charge density $\rho_\text{e,TIC}$ can be approximated as: 
\begin{eqnarray}
 \rho_\text{e,TIC} = -\frac{\varepsilon_0\varepsilon E_\text{pore}}{L} \Big( \frac{\Delta{n_\text{0,pore}}}{n_0} \Big),
&&\label{eq:four}
\end{eqnarray}
where $E_\text{pore}$, $\Delta n_\text{0,pore}$ and $L$ are the local electric field, salt concentration difference across the nanopore and pore length, respectively. Using Hall's form of access resistance, where the electric potential drops are analyzed by a circuit model~\cite{Hall1975}. The net resistance equals the sum of the pore resistance $R_\text{pore}=\frac{L}{\sigma_\text{c}\pi R^2}$ and access resistance  $R_\text{access}=\frac{1}{2\sigma_\text{c} R}$ at the junction of the nanopore and reservoirs.  Consequently, $E_\text{pore}$ can be expressed as a function of the external electric potential difference $\Delta \psi$ as:
\begin{eqnarray}
 E_\text{pore}=-\frac{2\Delta\psi}{\pi R+2L}.
&&\label{eq:five}
\end{eqnarray}
On the other hand, by assuming that $\rho_\text{e,TIC}$ is uniform and continuous at the junction between the nanopore and hemisphere regions along the axis, one can derive the product of the potential drop and concentration drop is constant for each region, yielding:   
\begin{eqnarray}
 \Delta{n_\text{0,pore}}=\frac{\pi L\Delta{n_0}}{8R+\pi L}.
&&\label{eq:six}
\end{eqnarray}

\begin{figure*}[hbt!]
	\includegraphics[width=\textwidth]{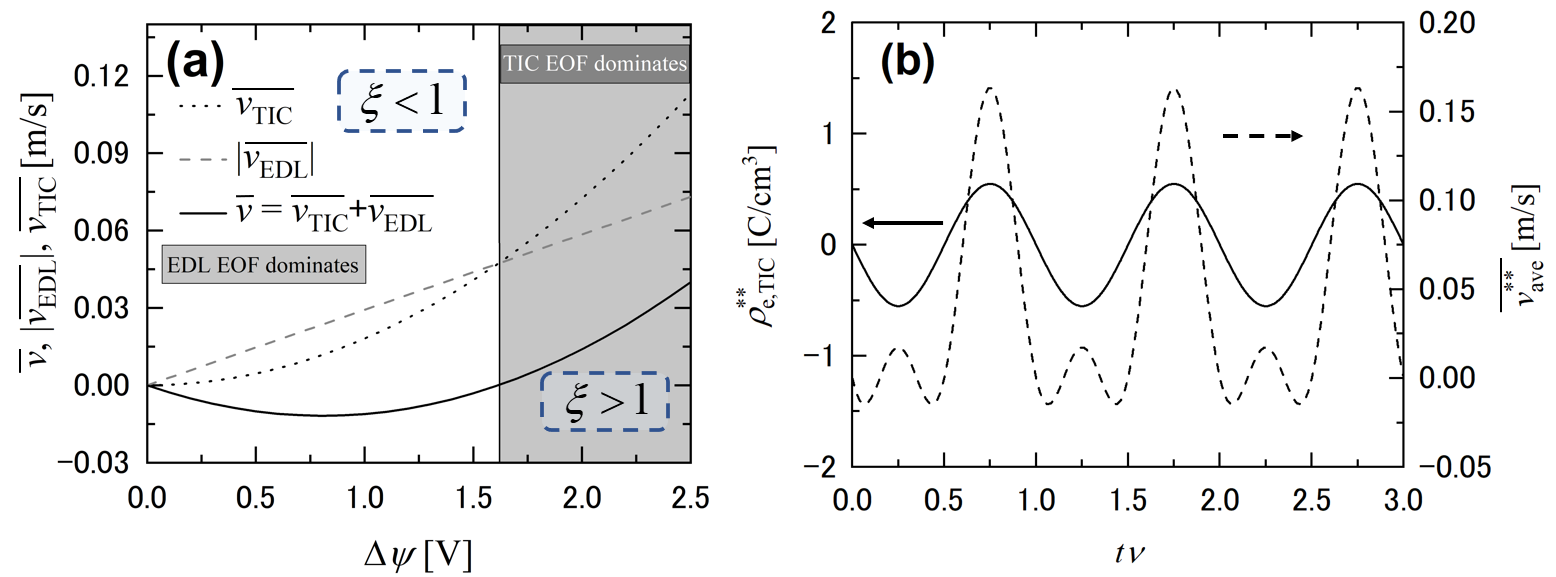}
	\caption{(a) Variations of steady-state electroosmotic mean velocity $\overline{v}$, and its velocity components $|\overline{v_{\text{EDL}}}|$ and $\overline{v_{\text{TIC}}}$ from EDL EOF and TIC EOF, respectively, as a function of applied electric field $\Delta\psi$ at $\Delta n_0 = 0.5 $ M. (b) Variations of AC induced charge density $\rho_\text{e,TIC}^{**}(t)$ (solid curve) and mean AC electroosmotic velocity $\overline{v^{**}}(t)$ (dashed curve) as functions of dimensionless time ($t\nu$) at $\Delta\psi$ = 2.5 V, $\Delta n_0 = 0.1$ M and $\nu$ = 100 kHz. ($n_0$ = 1 M, $\zeta$ = 1 mC/m$^2$, $L$ = 20 nm and $R$ = 2.5 nm.)   \label{fig3} }
\end{figure*}

The electroosmotic velocity in the axial direction $v_{z}$ can be described by the Navier-Stokes equation:
\begin{eqnarray}
 \eta\frac{1}{r}\frac{d}{dr}\Big(r\frac{dv_z(r)}{dr}\Big)+\rho_\text{e}E_\text{pore}=0,
&&\label{eq:seven}
\end{eqnarray}
where $\eta$ is the dynamic viscosity, which can be assumed as constant when viscoelectric effects are insignificant at low surface charge density~\cite{Hsu2017}. Eq.~(7) assumes the inertia force and pressure difference rising from the pore boundary effects at the inlet and outlet are small (which will be examined by numerical simulation for our system in the following section). The electroosmotic flow is driven by the electric force acting on both the charge of the EDL and TIC, $viz$ $v_{z}(r)=v_{z,\text{TIC}}(r)+v_{z,\text{EDL}}$ by considering linear superposition of these two effects, where $v_{z,\text{TIC}}(r)$ follows a parabolic distribution as:  
\begin{eqnarray}
 \begin{aligned}
 v_{z,\text{TIC}}(r)=- \frac{\varepsilon_0\varepsilon R^2E_\text{pore}^2}{4\eta L} \Big(\frac{\Delta{n_\text{0,pore}}}{n_0}\Big)\bigg[1-\Big(\frac{r}{R}\Big)^2\bigg], 
 &&\label{eq:eight}
 \end{aligned}
\end{eqnarray}
 and the mean TIC electroosmotic velocity $\overline{v_\text{TIC}}$ becomes:
\begin{eqnarray}
 \begin{aligned}
 \overline{v_\text{TIC}}=-\frac{\varepsilon_0\varepsilon R^2E_\text{pore}^2}{8\eta L} \Big(\frac{\Delta{n_\text{0,pore}}}{n_0}\Big).
 &&\label{eq:nine}
 \end{aligned}
\end{eqnarray}

To verify the analytical model, we conduct two-dimensional numerical simulations in cylindrical coordinates ($r$,$z$) using the coupled Poisson-Nernst-Planck, Navier-Stokes and continuity equations for incompressible fluid as below:
\begin{eqnarray}
 \nabla^2\psi=-\frac{\rho_\text{e}}{\varepsilon_0\varepsilon},
&&\label{eq:ten}
\end{eqnarray}
\begin{eqnarray}
 \frac{\partial n_i}{\partial t}=-\nabla\cdot\Big(-D_i\nabla n_i-\frac{D_iz_ie}{k_\text{B}T}\nabla\psi+n_i\textbf{v}\Big),
&&\label{eq:eleven}
\end{eqnarray}
\begin{eqnarray}
  \rho\bigg(\frac{\partial\textbf{v}}{\partial t}+\textbf{v}\cdot\nabla\textbf{v}\bigg)=-\nabla p+\eta\nabla^2\textbf{v}-\rho_\text{e}\nabla\psi,
&&\label{eq:twelve}
\end{eqnarray}
\begin{eqnarray}
 \nabla\cdot\textbf{v}=0.
&&\label{eq:thirteen}
\end{eqnarray}
In these expressions, $k_\text{B}$ is Boltzmann's constant, $T$ the temperature,  $e$ the unit charge, $n_i$ the ionic concentration, $D_i$ the ionic diffusivity and $z_i$ the ionic valence, where $i$ denotes  $+$ or $-$ for cations or anions, respectively. For simplicity, we consider a monovalent potassium chloride (KCl) aqueous solution, in which $z_+$=$-z_-$=1. In the modified Navier-Stokes equation considering an electric body force to the charged solution, \textbf{v} is the velocity vector and $p$ represents the pressure.  
We consider the surface of the nonconductive nanopore is ion-impermeable and non-slip. More simulation details can be found elsewhere~\cite{Hsu2018}. 

A comparison between the analytical solutions and numerical results at steady state is summarized in Fig.~2(a) and 2(b), showing close agreement for both $\rho_\text{e,TIC}$ and $\overline{v_\text{TIC}}$. 
Typically, the error percentage is less than 5$\%$ for $\rho_\text{e,TIC}$ and 10$\%$ for $v_\text{TIC}$. 
Note that, the simulation results show the same characteristics that $\rho_\text{e,TIC}\propto \Delta\psi\Delta n_0$ and $\overline{v_\text{TIC}}\propto (\Delta\psi)^2\Delta n_0$ following Eqs.~(4) and (9), yielding TIC EOF more dominant over EDL EOF at high applied electric fields.  
The simulated pressure difference across the nanopore due to the end effect is in the order of mPa within the investigated ranges of parameters, justifying the assuption in Eq.~7. In a similar verin, $v_{z,\text{EDL}}$ can be approximated by the Helmholtz-Smoulchowski equation as $\lambda_\text{D} << R$. For thin EDLs ($e.g.$, $\lambda_\text{D}\cong 0.3$ nm at $n_0$ = 1 M), the mean EDL electroosmotic velocity $\overline{v_{\text{EDL}}}$ = $v_{z,\text{EDL}}$ due to the plug flow behavior:
\begin{eqnarray}
 \begin{aligned}
 \overline{v_{\text{EDL}}} = v_{z,\text{EDL}}=-\frac{\varepsilon_0\varepsilon\zeta E_\text{pore}}{\eta}. 
 &&\label{eq:fourteen}
 \end{aligned}
\end{eqnarray}
 Therefore, the mean electroosmotic velocity $\overline{v}$ = $\overline{v_{\text{TIC}}}$ + $\overline{v_{\text{EDL}}}$ can be derived as: 
\begin{eqnarray}
 \begin{aligned}
 \overline{v}=-\frac{\varepsilon_0\varepsilon}{\eta} \bigg[\frac{R^2E_\text{pore}^2}{8L} \Big(\frac{\Delta{n_\text{0,pore}}}{n_0}\Big) +\zeta E_\text{pore} \bigg].
 &&\label{eq:fifteen}
 \end{aligned}
\end{eqnarray}

We define a dimensionless number $\xi$ representing the ratio of $\Delta\psi$ and $\zeta$ as a function of $R$, $L$ and $\Delta{n_\text{0,pore}}$ as:
 \begin{eqnarray}
 \begin{aligned}
 \xi=\frac{\Delta\psi R^2}{4L(\pi R+2L)\zeta}\bigg(\frac{\Delta{n_\text{0,pore}}}{n_0}\bigg).
 &&\label{eq:sixteen}
 \end{aligned}
\end{eqnarray}
where $\xi$ = 1, when $\overline{v}=0$. Hence, when $\xi << 1$ , the electroosmotic flow (EOF) from the EDL is dominant, whereas the TIC EOF governs the flow behavior at high applied voltages as $\xi  >> 1$. Figure~3a shows the variation of $\overline{v}$ and its velocity components with $\Delta\psi$.  Due to the nonlinear response of $\overline{v_{\text{TIC}}}$, the net velocity reverses at $\xi=1$ ($\Delta\psi\cong 1.6$ V), at which the contribution of TIC EOF surpasses that of EDL EOF.  

$\emph{Transient \textemdash}$ In the following, we consider transient variation of induced charge $\rho^*_\text{e,TIC}(t)$ at a fixed external potential difference $\Delta\psi$. Considering the case where $\varepsilon$ is constant and $\rho^*_\text{e,TIC}(t)$ is uniform in the nanopore, we derive a first order differential equation of $\rho^*_\text{e,TIC}$ according to Eqs.~(1) and (2):
\begin{eqnarray}
\begin{aligned}
 \frac{d\rho^*_\text{e,TIC}(t)}{dt}+\frac{\sigma_\text{c}}{\varepsilon_0\varepsilon} \rho^*_\text{e,TIC}(t) = -
 \frac{\sigma_\text{c} E_\text{pore}}{L}\Big( \frac{\Delta{n_\text{0,pore}}}{n_0}\Big). 
&&\label{eq:seventeen}
\end{aligned}
\end{eqnarray}
Eq.~(17) is then solved by an initial condition $\rho^*_\text{e,TIC}(0)=0$ considering no induced charge when the electric field is absent. We derive the transient variation of $\rho^*_\text{e,TIC}(t)$ as:
\begin{eqnarray}
 \rho^*_\text{e,TIC}(t)=-\frac{\varepsilon_0\varepsilon E_\text{pore}}{L} \Big( \frac{\Delta{n_\text{0,pore}}}{n_0}\Big)\Big(1-e^{-t/\tau}\Big),
&&\label{eq:eighteen}
\end{eqnarray}
and hence the transient mean TIC electroosmotic velocity $\overline{v_\text{TIC}^{*}}$ is derived by assuming a pseudo-steady-state response to $\rho^*_\text{e,TIC}(t)$ for the flow behavior:
\begin{eqnarray}
\overline{v_\text{TIC}^{*}}(t)=-\frac{\varepsilon_0\varepsilon R^2E_\text{pore}^2}{8\eta L} \Big(\frac{\Delta{n_\text{0,pore}}}{n_0}\Big)\Big(1-e^{-t/\tau}\Big),
&&\label{eq:nineteen}
\end{eqnarray}
where the characteristic time $\tau$ is (as $z_+$=$-z_-$=1):
\begin{eqnarray}
\tau=\frac{\varepsilon_0\varepsilon}{\sigma_\text{c}}=\frac{\varepsilon_0\varepsilon k_\text{B}T}{2e^2n_0\overline{D}}.
&&\label{eq:twenty}
\end{eqnarray}
Here, $\overline{D}$ is the arithmetic mean ionic diffusivity.
Fig.~2c shows good agreement between the analytical and numerical solutions of transient variations of $\rho^*_\text{e,TIC}(t)$ and $\overline{v_\text{TIC}^{*}}(t)$. $\tau$ is derived $\cong$ 46.3 ps for KCl solutions at $n_0$ = 1 M and $T$ = 298 K, where the charge induction process reaches steady state in around a hundred picoseconds.    

\begin{figure}[t!]
	\includegraphics[width=\columnwidth]{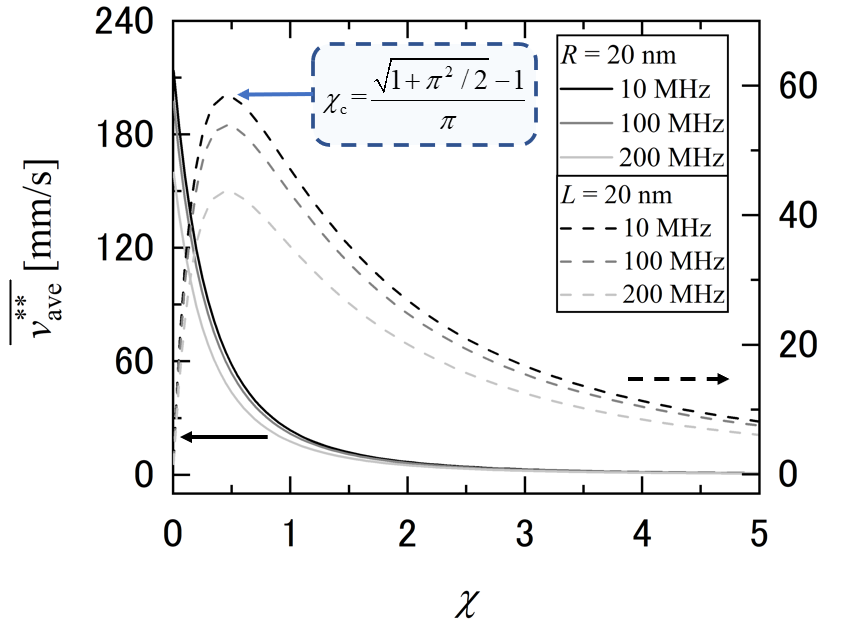}
	\caption{Variation of time-averaged mean velocity $\overline{v_\text{ave}^{**}}$, over each period $1/\nu$ under an AC sinusoidal field, as functions of the nanopore aspect ratio $\chi= L/D$ at different levels of frequency $\nu$. $\Delta\psi$ = 2.5 V and $\Delta n_0 = 0.1$ M. $L$ and $R$ are fixed at 20 nm for the solid curves and dashed curves, respectively. \label{fig4}}
\end{figure}

$\emph{AC fields \textemdash}$ Under an AC field, where the applied electric potential difference $\Delta\psi_\text{AC}(t)=\alpha\sin{(2\pi\nu t)}$ follows a sinusoidal function of an amplitude $\alpha$ and a frequency $\nu$, the AC induced charge density $\rho_\text{e,TIC}^{**}(t)$ follows Eq.~(17):
\begin{eqnarray}
 \frac{d\rho^{**}_\text{e,TIC}(t)}{dt}+\beta \rho^{**}_\text{e,TIC}(t) = 
 \gamma\Delta\psi_\text{AC}(t).
&&\label{eq:twentyone}
\end{eqnarray}
In these expressions, $\beta=1/\tau$, $\gamma=\frac{2\sigma_\text{c}}{(\pi R+2L)L}(\frac{\Delta{n_\text{0,pore}}}{n_0})$.
Using the initial condition $\rho^{**}_\text{e,TIC}(0)=0$, $\rho^{**}_\text{e,TIC}(t)$ can be solved as:
\begin{eqnarray}
  \rho_\text{e,TIC}^{**}(t)=\frac{\alpha\beta\gamma\sin{(\omega t)}-\alpha\gamma\omega[\cos{(\omega t)}-e^{-\beta t}]}{\beta^2+\omega^2 },
&&\label{eq:twentytwo}
\end{eqnarray}
where $\omega = 2\pi\nu$. Accordingly, the mean AC flow velocity $\overline{v^{**}}(t)$ contributed by both EDL EOF and TIC EOF, can be expressed as: 
\begin{eqnarray}
 \overline{v^{**}}(t)=\frac{2\Delta\psi_\text{AC}(t)}{\pi R+2L}\bigg(\frac{ \rho_\text{e,TIC}^{**}(t)R^2}{8\eta}+\frac{\varepsilon_0\varepsilon \zeta }{\eta}\bigg).
&&\label{eq:twentythree}
\end{eqnarray}
Typical behaviors of $\rho_\text{e,TIC}^{**}(t)$ and $\overline{v^{**}}(t)$ under an AC field are shown in Fig.~3b. Given that $\xi\cong 0.3$ in this case, the EDL EOF is dominant as $\rho_\text{e,TIC}^{**}(t)$ < 0, resulting in a negative velocity. Whereas, both the EDL EOF and TIC EOF are in the positive direction as $\rho_\text{e,TIC}^{**}(t)$ > 0 giving rise to obvious rectification of $\overline{v^{**}}(t)$ between different voltage directions. 

The time-averaged mean velocity $\overline{v_\text{ave}^{**}}$ during each period $1/\nu$ can be derived by integrating $\overline{v^{**}}(t)$ over time:  
\begin{eqnarray}
\begin{aligned}
 \overline{v_\text{ave}^{**}} & =\nu\int_0^{1/\nu} \overline{v^{**}}(t) dt \\ &= \frac{ \alpha^2\beta\gamma R^2}{8\eta(\beta^2+\omega^2)(\pi R+2L)}\neq f(\zeta) \\& \cong\frac{ \alpha^2\gamma R^2}{8\eta\beta(\pi R+2L)}, \text{when } \omega << \beta.
&&\label{eq:twentyfour}
\end{aligned}
\end{eqnarray}
Importantly, due to the net zero contribution from the EDL EOF for each cycle, $\overline{v_\text{ave}^{**}}$, which is proportional to $\alpha^2$, is not a function of $\zeta$ no matter the magnitude of $\xi$. The variation of $\overline{v_\text{ave}^{**}}$ as a function of the pore aspect ratio $\chi(=L/D)$, where $D = 2R$ is the pore diameter, is shown in Fig.~4. For a fixed $R$, $\overline{v_\text{ave}^{**}}$ monotonically increases for a shorter $L$ (solid curves). Conversely, $\overline{v_\text{ave}^{**}}$ experiences a maximum value (as the first derivative of $\overline{v_\text{ave}^{**}}$, with respect to $\chi$, is zero) when increasing $\chi$ at a fixed $L$, which occurs when $\chi$ is equivalent to a critical value $\chi_\text{c}$, and:
\begin{eqnarray}
\chi_\text{c}=\frac{\sqrt{1+\pi^2/2}-1}{\pi}\cong 0.457.
&&\label{eq:twentyfive}
\end{eqnarray}
This indicates that the maximum flow rate per unit area can be achieved using low aspect ratio pores whose length and radius are similar. At $n_0$ = 1M, $\beta/2\pi$ is approximately $\num{3.4e8}$, therefore $\overline{v_\text{ave}^{**}}$ is insensitive to the frequency when $\nu << 100$ MHz.

In conclusion, we analytically investigated TICEO for three cases: (i) at steady state, (ii) during transient variation at a constant applied voltage and (iii) during transient variation when a sinusoidal AC field is applied. We obtain a characteristic time $\tau$ (Eq.~20) for TIC phenomena. When the angular frequency of the AC field is much smaller than 1/$\tau$, the time-averaged mean flow velocity can be regarded as constant insensitive to the frequency. Furthermore, we define a dimensionless number $\xi$ to evaluate the dominance of TICEO (Eq.~16). When $\xi >> 1$, TIC EOF is dominant over EDL EOF, whereas the reverse is the case as $\xi << 1$. Finally, the optimized pore aspect ratio $\chi_\text{c}$ is determined, which shows that when $L$ $\cong$ 0.457$D$, the largest mean velocity under sinusoidal AC fields would be achieved. 
These results provide direct guidance to nanopore design for AC pumping and shed light on the principal mechanism of nonlinear electroosmosis in nanopores, paving the way toward a more precise flow control in multiple applications of nanopore technology.


\begin{acknowledgments}
The authors would like to thank Juan G. Santiago and Ali Mani at Stanford University as well as Jongyoon Han and Rohit Karnik at Massachusetts Institute of Technology for their comments.
This research was funded by the Japan Society for the Promotion of Science KAKENHI (Grants-in-Aid for Early Scientists 19K15600) and the University of Tokyo GAP Fund Program.
\end{acknowledgments}



\providecommand{\noopsort}[1]{}\providecommand{\singleletter}[1]{#1}%

\end{document}